\RequirePackage{fix-cm}
\documentclass[smallextended]{svjour3}       % onecolumn (second format)
\smartqed  % flush right qed marks, e.g. at end of proof
\usepackage{graphicx}
\usepackage{colortbl}
\usepackage{mathptmx}      % use Times fonts if available on your TeX system
\begin{document}

\title{Controlled Remote State Preparation via General Pure Three-Qubit State %Insert your title here%\thanks{Grants or other notes
}

\author{Zhi-Hua Zhang \and Jun Zheng \and Lan Shu  
}


\institute{Zhi-Hua Zhang \and
       Lan Shu  \at
              School of Mathematical Sciences, University of Electronic Science and Technology of China, Chendu 611731, Sichuan Province, China \\
              \email{zhihuamath@aliyun.com}    \\
           \and
           Jun Zheng \at
           Basic Course Department, Emei Campus, Southwest Jiaotong University, Sichuan, Emei 614202, P. R. China\\
}

\date{Received: date / Accepted: date}

\maketitle

\begin{abstract}
The protocols for controlled remote state preparation of a single qubit and a general two-qubit state
are presented in this paper. The general pure three-qubit states are chosen as shared quantum channel, which are not LOCC equivalent to the
mostly used GHZ-state. It is the first time to introduce
general pure three-qubit states to complete remote state preparation. The probability of successful
preparation is presented. Moreover, in some special cases, the successful probability could reach unit.

%Insert your abstract here. Include keywords, PACS and mathematical
%subject classification numbers as needed.
\end{abstract}

\section{Introduction}
\label{intro}
%Your text comes here. Separate text sections with
Quantum teleportation (QT for short) is the first quantum information
processing protocol presented by Bennett \emph{et al.} \cite{Ben1}
to achieve the transmission of information contained in quantum state
determinately. Many theoretical schemes have been proposed later
\cite{Lee1,Kim,Zhou,Wang1,van}. It has also been realized experimentally
\cite{Bou,Bra,Fur,Jin,Hua,Noe,Nil,Fri}.
Latter, to save resource needed in the process of information transmission,
Lo put forward a scheme for remote preparation of quantum state
(RSP for short) \cite{Lo}. Compared with QT, in RSP the sender does not own the
particle itself but owns all the classical information of the state he or she
wants to prepare for the receiver, who is located separately from the
sender. The resource consumption is reduced greatly in RSP, as the sender do not need to prepare the state beforehand. The RSP has already attracted
many attentions. A number of RSP protocols were presented, such as RSP with or
without oblivious conditions, optimal RSP, RSP using noisy channel, low-entanglement RSP,
continuous variable RSP and so on \cite{Ben2,Dev,Guo,Zeng,Shi,Berry,Leung,Hayashi,Yu,Kur,Lee2}.
Experimental realization was also proved \cite{Peng,Bar}.

In RSP protocols, all the classical information is distributed to one sender, which may lead to information leakage if the sender is not honest. In order to improve the security of remote state preparation, controllers are introduced, which is the so called controlled remote state preparation (CRSP for short), and it has drawn the attention of many researchers. In contrast to the usual RSP, the CRSP needs to incorporate a controller. The information could be transmitted if and only if both the sender and receiver cooperate with the controller or supervisor. CRSP for an arbitrary qubit has been presented in a network via many agents \cite{WangZY1}. A two-qubit state CRSP with multi-controllers using two non-maximally GHZ states as shared channel is shown in \cite{HouK}. CRSP with two receivers via asymmetric channel \cite{WangZY2},
using POVM are presented \cite{WangZY3,LiZ}. The five-qubit Brown state as quantum
channel to realize the CRSP of three-qubit state is elaborated in \cite{ChenXB}.
Most of the existing schemes chose to use the GHZ-type state, W-type state, Bell state or
the composite of these states as the shared quantum channel. However in this paper, we choose
the general pure three-qubit state as quantum channel, which is not LOCC equivalent to the GHZ state.
And for some special cases, the probability for successful CRSP can reach unit.

In \cite{AAcin}, the authors proved that for any pure three-qubit state, the existence
of local base, which allows one to express a pure three-qubit state in a unique form using a
set of five orthogonal state. It is the called generalised Schmidt-Decomposition
for three-qubit state. Using the generalised Schmidt-Decomposition, Gao \emph{et al.} \cite{GaoT}
proposed a controlled teleportation protocol for an unknown qubit and gave analytic
expressions for the maximal successful probabilities. They also gave an explicit
expression for the pure three-qubit state with unit probability of controlled teleportation \cite{GaoT}. Motivated by the ideas of the two papers, we try to investigate the controlled remote state preparation using the general pure three-qubit states and their generalised Schmidt-Decomposition.

The paper is arranged as follows. In Sec. 2, the CRSP for an arbitrary qubit is elucidated in detail.
We find that the successful probability is the same as that of controlled teleportation for qubits
with real coefficients. In Sec. 3, the CRSP for a general two-qubit state is expounded. For two-qubit state with four real coefficients. The corresponding successful probability is the same as that of controlled teleportation of a qubit. In Sec. 4, we conclude the paper.

\section{CRSP for an arbitrary qubit}

Suppose that three separated parties Alice, Bob and Charlie share a general pure three-qubit particle
$|\Phi\rangle_{cab}$, the particle $a$ belongs to Alice, $b$ to Bob and $c$ to Charlie, respectively.
The distribution of the three particles are sketched in Fig.1. 
\begin{figure}
 \includegraphics[width=0.35\textwidth]{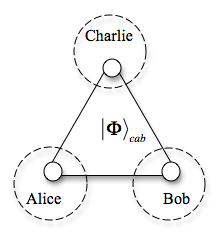}
\caption{Particle distribution in one qubit CRSP}
\label{fig1}       % Give a unique label
\end{figure}
In figure 1, the small circles represent the particles, the solid line between two circles means that the corresponding two particles are related to each other by quantum correlation. According to \cite{AAcin}, the general pure three qubit state has a unique generalised
Schmidt-Decomposition in the form
\begin{equation}
|\Phi\rangle_{cab}
=(a_{0}|000\rangle+a_{1}e^{i\mu}|100\rangle+a_{2}|101\rangle+a_{3}|110\rangle+a_{4}|111\rangle)_{cab},
\end{equation}
where $a_{i}\geq 0$ for $i=0,\cdots, 4$, $0\leq \mu \leq \pi$, $\sum_{i=0}^{4}a_{i}^{2}=1$.
The $a_{i}$ and $\mu$ in Eq.(1) are decided uniquely with respect to a chosen general pure
three qubit state.

Now Alice wants to send the information of a general qubit
$$|\varphi\rangle=\alpha|0\rangle+\beta|1\rangle, \quad |\alpha|^{2}+|\beta|^{2}=1$$
to the remote receiver Bob under the control of Charlie. Alice possesses the classical information of
this qubit, i.e. the information of $\alpha$ and $\beta$, but does not have the particle itself.
Next, we make three steps to complete the CRSP for $|\varphi\rangle$.

\emph{Step 1} The controller Charlie firstly makes a single qubit measurement under the base
\begin{equation}
|\varepsilon_{c}^{0}\rangle=\cos\frac{\theta}{2}|0\rangle+e^{i\eta}\sin\frac{\theta}{2}|1\rangle, \quad
|\varepsilon_{c}^{1}\rangle=\sin\frac{\theta}{2}|0\rangle-e^{i\eta}\cos\frac{\theta}{2}|1\rangle,
\end{equation}
where $\theta\in [0,\pi]$, $\eta\in [0,2\pi]$. The choice of $\theta$ and $\eta$ could be flexible according to the need of the controller. If $\theta=\pi$ and $\eta=0$, $|\varepsilon_{c}^{0}\rangle$ and $|\varepsilon_{c}^{1}\rangle$ will be the $|\pm\rangle$ base. Then Charlie broadcasts his measurement outcomes publicly to Alice and Bob using one classical bit. Using Eq.(2), the quantum channel can be rewritten as
\begin{equation}
|\Phi\rangle_{cab}=\sqrt{p_{0}}|\varepsilon_{c}^{0}\rangle|\Omega_{0}\rangle_{ab}
+\sqrt{p_{1}}|\varepsilon_{c}^{1}\rangle|\Omega_{1}\rangle_{ab},
\end{equation}
where
$$p_{0}=\sin^{2}\frac{\theta}{2}+a_{0}^{2}\cos\theta+a_{0}a_{1}\cos(\mu-\eta)\sin\theta,$$
$$p_{1}=\cos^{2}\frac{\theta}{2}-a_{0}^{2}\cos\theta-a_{0}a_{1}\cos(\mu-\eta)\sin\theta,$$
\begin{eqnarray*}
|\Omega_{0}\rangle_{ab}&&=\frac{1}{\sqrt{p_{0}}}\bigg\{[a_{0}\cos\frac{\theta}{2}
+a_{1}e^{i(\mu-\eta)}\sin\frac{\theta}{2}]|00\rangle
\nonumber\\
&&\quad +e^{-i\eta}\sin\frac{\theta}{2}[a_{2}|01\rangle
+a_{3}|10\rangle+a_{4}|11\rangle]\bigg\}_{ab}
\end{eqnarray*}
\begin{eqnarray*}
|\Omega_{1}\rangle_{ab}&&=\frac{1}{\sqrt{p_{1}}}\bigg \{[a_{0}\sin\frac{\theta}{2}
-a_{1}e^{i(\mu-\eta)}\cos\frac{\theta}{2}]|00\rangle
\nonumber\\
&&\quad-e^{-i\eta}\cos\frac{\theta}{2}[a_{2}|01\rangle
+a_{3}|10\rangle+a_{4}|11\rangle] \bigg\}_{ab}
\end{eqnarray*}
If the result of Charlie's measurement is $0$, the whole system collapses to $|\Omega_{0}\rangle_{ab}$
with probability $p_{0}$ while collapses to $|\Omega_{1}\rangle_{ab}$ with probability $p_{1}$ for the
result $1$. To ensure that the particle $c$ entangles with the whole system, we assume that $a_{0}>0$ and
$a_{2}, a_{3}, a_{4}$ are not equal to $0$ at the same time. This is equivalent to $p_{0}>0$
and $p_{1}>0$ at the same time.

Note that \emph{Step 1} is actually similar to that of controlled
teleportation in \cite{GaoT}. We arrange it here to keep the integrity of the paper. More detailed
calculation can be found in \cite{GaoT}.

\emph{Step 2} Without loss of generality, we assume that the result of Charlie's measurement is $0$.
Then the whole system collapse to $|\Omega_{0}\rangle_{ab}$. Using the Schmidt-Decomposition of two-qubit system, there exists bases $\{|0^{'}\rangle, |1^{'}\rangle\}_{a}$ and $\{|0^{'}\rangle, |1^{'}\rangle\}_{b}$
for particle $a$ and $b$ respectively, such that $|\Omega_{0}\rangle_{ab}$ can be expressed as
\begin{equation}
|\Omega_{0}\rangle_{ab}=(\sqrt{\lambda_{00}}|0^{'}0^{'}\rangle+\sqrt{\lambda_{01}}|1^{'}1^{'}\rangle)_{ab},
\end{equation}
where $\lambda_{00}=(1-\sqrt{1-C_{0}^{2}})/2$, $\lambda_{01}=(1+\sqrt{1-C_{1}^{2}})/2$ in \cite{GaoT}.
On receiving the result of Charlie's measurement, the sender Alice prepares a projective measurement
utilizing the classical information of $|\varphi\rangle$ in the following form:
\begin{equation}
\left(
  \begin{array}{c}
    |\mu_{0}\rangle \\
    |\mu_{1}\rangle \\
  \end{array}
\right)_{a}
=\left(
   \begin{array}{cc}
    \alpha & \beta \\
     \beta^{*} & -\alpha^{*} \\
   \end{array}
 \right)\left(
          \begin{array}{c}
            |0^{'}\rangle \\
            |1^{'}\rangle \\
          \end{array}
        \right)_{a}.
\end{equation}
Then $|\Omega_{0}\rangle_{ab}$ could be reexpressed as
\begin{equation}
|\Omega_{0}\rangle_{ab}=|\mu_{0}\rangle_{a}(\sqrt{\lambda_{00}}\alpha^{*}|0^{'}\rangle
+\sqrt{\lambda_{01}}\beta^{*}|1^{'}\rangle)_{b}
+|\mu_{1}\rangle_{a}(\sqrt{\lambda_{00}}\beta|0^{'}\rangle-\sqrt{\lambda_{01}}\alpha|1^{'}\rangle)_{b}.
\end{equation}
Next we first discuss the case for real coefficients, i.e. $\alpha, \beta$ are real. Then Eq.(6) will be
\begin{eqnarray*}
|\Omega_{0}\rangle_{ab}=|\mu_{0}\rangle_{a}(\sqrt{\lambda_{00}}\alpha|0^{'}\rangle
+\sqrt{\lambda_{01}}\beta|1^{'}\rangle)_{b}
+|\mu_{1}\rangle_{a}(\sqrt{\lambda_{00}}\beta|0^{'}\rangle-\sqrt{\lambda_{01}}\alpha|1^{'}\rangle)_{b}.
\end{eqnarray*}
Alice measures her qubit under base $\{|\mu_{0}\rangle, |\mu_{1}\rangle\}_{a}$ and gets the outcome $0$
and $1$ with probability $\lambda_{00}\alpha^{2}+\lambda_{01}\beta^{2}$ and
$\lambda_{00}\beta^{2}+\lambda_{01}\alpha^{2}$ respectively. And Alice sends her measurement result to
Bob by 1 classical bit. The receiver Bob's system will collapse to
\begin{equation}
|\xi_{0}\rangle_{b}=\frac{1}{\sqrt{\lambda_{00}\alpha^{2}
+\lambda_{01}\beta^{2}}}(\sqrt{\lambda_{00}}\alpha|0^{'}\rangle+\sqrt{\lambda_{01}}\beta|1^{'}\rangle)_{b},
\end{equation}
\begin{equation}
|\xi_{1}\rangle_{b}=\frac{1}{\sqrt{\lambda_{00}\beta^{2}
+\lambda_{01}\alpha^{2}}}(\sqrt{\lambda_{00}}\beta|0^{'}\rangle-\sqrt{\lambda_{01}}\alpha|1^{'}\rangle)_{b}
\end{equation}
respectively.

\emph{Step 3} We assume that Alice's measurement result is 0. Now according to Charlie and Alice's result,
Bob wants to recovery the state $|\varphi\rangle$ on his side. Bob needs to introduce an auxiliary particle
in initial state $|0\rangle_{b^{'}}$, then he makes a unitary operation $U_{bb^{'}}^{0}$ on his particle
$b$ and the auxiliary particle $b^{'}$, and his state changes to $|\omega^{0}\rangle_{bb^{'}}$, where 
$$
U_{bb^{'}}^{0}=\left(
  \begin{array}{cccc}
    1 & 0 & 0 & 0 \\
    0 & 1 & 0 & 0 \\
    0 & 0 &  \sqrt{\frac{\lambda_{00}}{\lambda_{01}}} & \sqrt{1-\frac{\lambda_{00}}{\lambda_{01}}} \\
    0 & 0 & -\sqrt{1-\frac{\lambda_{00}}{\lambda_{01}}} & \sqrt{\frac{\lambda_{00}}{\lambda_{01}}} \\
  \end{array}
\right),$$
\begin{eqnarray*}
|\omega^{0}\rangle_{bb^{'}}&=&U_{bb^{'}}^{0}|\xi_{0}\rangle_{b}|0\rangle_{b^{'}}
\nonumber\\&=&\frac{1}{\sqrt{\lambda_{00}\alpha^{2}+\lambda_{01}\beta^{2}}}
\left[\sqrt{\lambda_{00}}(\alpha|0^{'}0\rangle+\beta|1^{'}0\rangle)
+\sqrt{\lambda_{01}-\lambda_{00}}\beta|1^{'}1\rangle\right]_{bb^{'}}.
\end{eqnarray*}
After the unitary operation, Bob makes a measurement on his auxiliary particle $b^{'}$ under the base
$\{|0\rangle, |1\rangle\}_{b^{'}}$. The probability for Bob to get measurement result $0$ is
$\lambda_{00}/(\lambda_{00}\alpha^{2}+\lambda_{01}\beta^{2})$, and he can
recovery state $|\varphi\rangle$ successfully. But if the result is $1$, the scheme fails.

Similarly, if Alice's measurement result is 1, Bob also introduces an auxiliary particle in initial
state $|0\rangle_{b^{'}}$. But the unitary operation is $U_{bb^{'}}^{1}$, and the system after the
unitary operation is $|\omega^{1}\rangle_{bb^{'}}$, where
$$
U_{bb^{'}}^{1}=\left(
  \begin{array}{cccc}
    0 & 0 & 1 & 0 \\
    0 & 1 & 0 & 0 \\
   -\sqrt{\frac{\lambda_{00}}{\lambda_{01}}} & 0 &  0 & \sqrt{1-\frac{\lambda_{00}}{\lambda_{01}}} \\
    \sqrt{1-\frac{\lambda_{00}}{\lambda_{01}}} & 0 & 0 & \sqrt{\frac{\lambda_{00}}{\lambda_{01}}}\\
  \end{array}
\right),$$
\begin{eqnarray*}
|\omega^{1}\rangle_{bb^{'}}&=&U_{bb^{'}}^{1}|\xi_{1}\rangle_{b}|0\rangle_{b^{'}}
\nonumber\\&=&\frac{1}{\sqrt{\lambda_{00}\beta^{2}+\lambda_{01}\alpha^{2}}}
\left[\sqrt{\lambda_{00}}(\alpha|0^{'}0\rangle+\beta|1^{'}0\rangle)
-\sqrt{\lambda_{01}-\lambda_{00}}\alpha|1^{'}1\rangle\right]_{_{bb^{'}}}.
\end{eqnarray*}
The probability for Bob to successfully reconstruct the state $|\varphi\rangle$
is $\lambda_{00}/(\lambda_{00}\beta^{2}+\lambda_{01}\alpha^{2})$.

Combining the process
of \emph{Step 1} and \emph{Step 2}, when the controller Charlie's measurement result is 0,
the receiver Bob can reconstruct the qubit $|\varphi\rangle$ with probability
$$p_{0}(\lambda_{00}\alpha^{2}+\lambda_{01}\beta^{2})\frac{\lambda_{00}}{\lambda_{00}\alpha^{2}+\lambda_{01}\beta^{2}}
+p_{0}(\lambda_{00}\beta^{2}+\lambda_{01}\alpha^{2})\frac{\lambda_{00}}{\lambda_{00}\beta^{2}+\lambda_{01}\alpha^{2}}
=2p_{0}\lambda_{00}.$$

Similarly, if Charlie's measurement result is 1 with probability $p_{1}$, the whole system
collapses to $|\Omega_{1}\rangle_{ab}$. And there are bases $\{|\overline{0}\rangle, |\overline{1}\rangle\}_{a}$
and $\{|\overline{0}\rangle, |\overline{1}\rangle\}_{b}$ for Alice and Bob's systems (\cite{GaoT} for reference),
so that the Schmidt-Decomposition for $|\Omega_{1}\rangle_{ab}$ is
$$|\Omega_{1}\rangle_{ab}=(\sqrt{\lambda_{10}}|\overline{0} \, \overline{0}\rangle
+\sqrt{\lambda_{11}}|\overline{1} \, \overline{1}\rangle)_{ab}.$$
Then continuing to use the last 2 steps as those in Charlie’s measurement result is 0, we can get that the successful probability for Bob to
produce the desired state is $2p_{1}\lambda_{10}$.

As a result, for the real case, Alice can prepare the qubit $|\varphi\rangle$ at Bob's position under the control of Charlie with probability $2(p_{0}\lambda_{00}+p_{1}\lambda_{10})$, which is the same as that of controlled teleportation in \cite{GaoT}. But the consumption of classical bits is reduced to 2 cbits for the whole process.

Next we discuss the case for complex coefficients. \emph{Step 1} is the same as that of real case.
In \emph{Step 2}, if Alice's measurement result is 0, referring to Eq. (6), the remote state preparation fails. When Alice
gets the result 1 with probability $\lambda_{00}|\beta|^{2}+\lambda_{01}|\alpha|^{2}$, the whole system
collapses to
$$\frac{1}{\sqrt{\lambda_{00}|\beta|^{2}
+\lambda_{01}|\alpha|^{2}}}(\sqrt{\lambda_{00}}\beta|0^{'}\rangle-\sqrt{\lambda_{01}}\alpha|1^{'}\rangle).$$
Then \emph{Step 3} is the same as that of the real case. The whole successful probability is
\begin{eqnarray*}
&&p_{0}(\lambda_{00}|\beta|^{2}+\lambda_{01}|\alpha|^{2})\frac{\lambda_{00}}{\lambda_{00}|\beta|^{2}+\lambda_{01}|\alpha|^{2}}
+p_{1}(\lambda_{10}|\beta|^{2}+\lambda_{11}|\alpha|^{2})\frac{\lambda_{10}}{\lambda_{10}|\beta|^{2}+\lambda_{11}|\alpha|^{2}}\nonumber\\
&&\quad=p_{0}\lambda_{00}+p_{1}\lambda_{10},
\end{eqnarray*}
which is half of the real case.

According to the discussion of \cite{GaoT}, the maximally probability for controlled teleportation
will reach unit if and only if the shared channel is
$$a_{0}|000\rangle+a_{1}|100\rangle+\frac{1}{\sqrt{2}}|111\rangle, \quad a_{0}>0, \quad a_{1}\geq 0,
\quad a_{0}^{2}+a_{1}^{2}=\frac{1}{2}.$$
As for the controlled remote state preparation for a qubit using the above channel, the
successful probability can also reach one for the real case, and $1/2$ for the complex case.

\section{CRSP for a two-qubit state}
In the CRSP for a two-qubit state, there are also three parties Alice, Bob and Charlie. They share a
quantum channel which is the composite of $|\Phi\rangle_{cab}$ and the Bell state, the distribution of particles in the shared quantum channel is displayed in Fig.2, the meaning of symbols is the same as in Fig.1.
\begin{eqnarray*}
&&|\Phi\rangle_{cab}|\phi^{+}\rangle_{a^{'}b^{'}}
\nonumber\\&=&(a_{0}|000\rangle+a_{1}e^{i\mu}|100\rangle+a_{2}|101\rangle+a_{3}|110\rangle+a_{4}|111\rangle)_{cab}
\frac{1}{\sqrt{2}}(|00\rangle+|11\rangle)_{a^{'}b^{'}},
\end{eqnarray*}
the particle $c$ belongs to Charlie, $a,a^{'}$ to Alice and $b,b^{'}$ to Bob. Now the sender Alice
possesses the classical information of a general two qubit state $|\varphi\rangle$,
$$|\varphi\rangle=\alpha|00\rangle+\beta|01\rangle+\gamma|10\rangle+\delta|11\rangle, \quad
|\alpha|^{2}+|\beta|^{2}+|\gamma|^{2}+|\delta|^{2}=1,$$
she wants to prepare the state
at the position of a distant receiver Bob with the help of a controller Charlie. Like CRSP in Section 2, there are three
steps to complete this task.
\begin{figure}
 \includegraphics[width=0.45\textwidth]{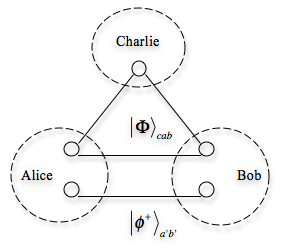}
\caption{Particle distribution in two-qubit CRSP}
\label{fig2}       % Give a unique label
\end{figure}

\emph{Step 1} This step is the same as that of \emph{Step 1} in Section 2. Charlie makes a projective measurement
$\{|\varepsilon_{c}^{0}\rangle, |\varepsilon_{c}^{1}\rangle\}$ on his
particle $c$, and gets the measurement result 0 and 1 with probability $p_{0}$ and $p_{1}$ respectively.
The whole system collapses to $|\Omega_{0}\rangle_{ab}|\phi^{+}\rangle_{a^{'}b^{'}}$ and
$|\Omega_{1}\rangle_{ab}|\phi^{+}\rangle_{a^{'}b^{'}}$ respectively. He broadcast his measurement result using 1 cbits.

\emph{Step 2} We assume Charlie's measurement result is 0 in \emph{Step 1}. Then the system state
after his measurement is $|\Omega_{0}\rangle_{ab}|\phi^{+}\rangle_{a^{'}b^{'}}$. Utilizing Schmidt-Decomposition \cite{GaoT},
there exists bases $\{|0^{'}\rangle, |1^{'}\rangle\}_{a}$ and $\{|0^{'}\rangle, |1^{'}\rangle\}_{b}$ such that
\begin{eqnarray*}
&&|\Omega_{0}\rangle_{ab}|\phi^{+}\rangle_{a^{'}b^{'}}
\nonumber\\&=&\frac{1}{\sqrt{2}}(\sqrt{\lambda_{00}}|0^{'}0^{'}\rangle
+\sqrt{\lambda_{01}}|1^{'}1^{'}\rangle)_{ab}(|00\rangle+|11\rangle)_{a^{'}b^{'}}
\nonumber\\&=&\frac{1}{\sqrt{2}}\left[\sqrt{\lambda_{00}}|0^{'}00^{'}0\rangle+\sqrt{\lambda_{00}}|0^{'}10^{'}1\rangle
+\sqrt{\lambda_{01}}|1^{'}01^{'}0\rangle+\sqrt{\lambda_{01}}|1^{'}11^{'}1\rangle\right]_{aa^{'}bb^{'}}.
\end{eqnarray*}
Next we first discuss the case in which all the coefficients are real. According to her knowledge of the two-qubit state $|\varphi\rangle$,
Alice constructs the measurement basis $\{|\mu_{0}\rangle, |\mu_{1}\rangle, |\mu_{2}\rangle, |\mu_{3}\rangle\}_{aa^{'}}$,
$$\left(
  \begin{array}{c}
    |\mu_{0}\rangle \\
    |\mu_{1}\rangle \\
    |\mu_{2}\rangle \\
    |\mu_{3}\rangle \\
  \end{array}
\right)_{aa^{'}}=\left(
          \begin{array}{cccc}
            \alpha & \beta   & \gamma  & \delta \\
            \beta  & -\alpha & -\delta & \gamma \\
            \gamma & \delta  & -\alpha & -\beta \\
            \delta & -\gamma & \beta   & -\alpha \\
          \end{array}
        \right)\left(
                 \begin{array}{c}
                   |0^{'}0\rangle \\
                   |0^{'}1\rangle \\
                   |1^{'}0\rangle \\
                   |1^{'}1\rangle \\
                 \end{array}
               \right)_{aa^{'}},
$$
Then the system for Alice and Bob can be rewritten as
\begin{eqnarray}
&&|\Omega_{0}\rangle_{ab}|\phi^{+}\rangle_{a^{'}b^{'}}
\nonumber\\&=&\frac{1}{\sqrt{2}}\bigg \{|\mu_{0}\rangle[\sqrt{\lambda_{00}}(\alpha|0^{'}0\rangle
+\beta|0^{'}1\rangle)+\sqrt{\lambda_{01}}(\gamma|1^{'}0\rangle
+\delta|1^{'}1\rangle)]
\nonumber\\&&+|\mu_{1}\rangle[\sqrt{\lambda_{00}}(\beta|0^{'}0\rangle
-\alpha|0^{'}1\rangle)-\sqrt{\lambda_{01}}(\delta|1^{'}0\rangle
-\gamma|1^{'}1\rangle)]
\nonumber\\&&+|\mu_{2}\rangle[\sqrt{\lambda_{00}}(\gamma|0^{'}0\rangle
+\delta|0^{'}1\rangle)-\sqrt{\lambda_{01}}(\alpha|1^{'}0\rangle
+\beta|1^{'}1\rangle)]
\nonumber\\&&+|\mu_{3}\rangle[\sqrt{\lambda_{00}}(\delta|0^{'}0\rangle
-\gamma|0^{'}1\rangle)+\sqrt{\lambda_{01}}(\beta|1^{'}0\rangle
-\alpha|1^{'}1\rangle)]\bigg\}_{aa^{'}bb^{'}}.
\end{eqnarray}
Thus Alice can get result 0 or 1 with probability
$[\lambda_{00}(\alpha^{2}+\beta^{2})+\lambda_{01}(\gamma^{2}+\delta^{2})]/2$, respectively, and result 3 or 4 with probability
$[\lambda_{00}(\gamma^{2}+\delta^{2})+\lambda_{01}(\alpha^{2}+\beta^{2}))]/2$.
The system state after Alice's measurement is
$$|\xi_{0}\rangle_{bb^{'}}=\frac{\sqrt{\lambda_{00}}(\alpha|0^{'}0\rangle
+\beta|0^{'}1\rangle)+ \sqrt{\lambda_{01}}(\gamma|1^{'}0\rangle
+\delta|1^{'}1\rangle)}{\sqrt{\lambda_{00}(\alpha^{2}+\beta^{2})
+\lambda_{01}(\gamma^{2}+\delta^{2})}},$$

$$|\xi_{1}\rangle_{bb^{'}}=\frac{\sqrt{\lambda_{00}}(\beta|0^{'}0\rangle
-\alpha|0^{'}1\rangle)- \sqrt{\lambda_{01}}(\delta|1^{'}0\rangle
-\gamma|1^{'}1\rangle)}{\sqrt{\lambda_{00}(\alpha^{2}+\beta^{2})
+\lambda_{01}(\gamma^{2}+\delta^{2})}},$$

$$|\xi_{2}\rangle_{bb^{'}}=\frac{\sqrt{\lambda_{00}}(\gamma|0^{'}0\rangle
+\delta|0^{'}1\rangle)- \sqrt{\lambda_{01}}(\alpha|1^{'}0\rangle
+\beta|1^{'}1\rangle)}{\sqrt{\lambda_{00}(\gamma^{2}+\delta^{2})
+\lambda_{01}(\alpha^{2}+\beta^{2})}},$$

$$|\xi_{3}\rangle_{bb^{'}}=\frac{\sqrt{\lambda_{00}}(\delta|0^{'}0\rangle
-\gamma|0^{'}1\rangle)+\sqrt{\lambda_{01}}(\beta|1^{'}0\rangle
+\alpha|1^{'}1\rangle)}{\sqrt{\lambda_{00}(\gamma^{2}+\delta^{2})
+\lambda_{01}(\alpha^{2}+\beta^{2})}}$$
with respective to the result 0, 1, 2, 3. Alice then broadcasts her measurement result to Bob
using 2 cbits.

\emph{Step 3} Assume that the measurement result of Alice is 0 in \emph{Step 2}. Then according to
the result, Bob introduces an auxiliary particle $b_{a}$ in the initial state
$|0\rangle_{b_{a}}$, and makes unitary operation $U_{bb^{'}b_{a}}^{0}$ on his particles, where
\begin{eqnarray*}
U_{bb^{'}b_{a}}^{0}=\left(
  \begin{array}{cc}
    I_{4} & 0   \\
        0 & U_{0} \\
  \end{array}
\right),
\end{eqnarray*}
here $I_{4}$ is the $4\times 4$ identity matrix and
\begin{eqnarray*}
U_{0}=\left(
        \begin{array}{cccc}
        \sqrt{\frac{\lambda_{00}}{\lambda_{01}}} & 0 & 0 & \sqrt{1-\frac{\lambda_{00}}{\lambda_{01}}} \\
          0 & -\sqrt{\frac{\lambda_{00}}{\lambda_{01}}} & \sqrt{1-\frac{\lambda_{00}}{\lambda_{01}}}  & 0 \\
          0 & \sqrt{1-\frac{\lambda_{00}}{\lambda_{01}}}  & \sqrt{\frac{\lambda_{00}}{\lambda_{01}}} & 0 \\
          \sqrt{1-\frac{\lambda_{00}}{\lambda_{01}}} & 0 & 0 & -\sqrt{\frac{\lambda_{00}}{\lambda_{01}}} \\
        \end{array}
      \right).
\end{eqnarray*}
The state after Bob performing the unitary operation is
\begin{eqnarray*}
&&U_{bb^{'}b_{a}}^{0}|\xi_{0}\rangle_{bb^{'}}|0\rangle_{b_{a}}
\nonumber\\&=&\frac{\sqrt{\lambda_{00}}(\alpha|0^{'}0\rangle+\beta|0^{'}1\rangle+\gamma|1^{'}0\rangle
+\delta|1^{'}1\rangle)|0\rangle+\sqrt{\lambda_{01}-\lambda_{00}}(\gamma|1^{'}0\rangle
+\delta|1^{'}1\rangle)|1\rangle}
{\sqrt{\lambda_{00}(\alpha^{2}+\beta^{2})+\lambda_{01}(\gamma^{2}+\delta^{2})}}.
\end{eqnarray*}
Thereafter, Bob makes a projective measurement on his auxiliary particles under basis
$\{|0\rangle, |1\rangle\}_{b_{a}}$.
He can get result 0 with probability
$\lambda_{00}/(\lambda_{00}(\alpha^{2}+\beta^{2})+\lambda_{01}(\gamma^{2}+\delta^{2}))$. As for the
other three cases, Bob can successfully reconstruct the desired two qubit state with
probability $\lambda_{00}/(\lambda_{00}(\alpha^{2}+\beta^{2})+\lambda_{01}(\gamma^{2}+\delta^{2}))$,
$\lambda_{00}/(\lambda_{00}(\gamma^{2}+\delta^{2})+\lambda_{01}(\alpha^{2}+\beta^{2}))$, and
$\lambda_{00}/(\lambda_{00}(\gamma^{2}+\delta^{2})+\lambda_{01}(\alpha^{2}+\beta^{2}))$.

Similarly, in the real case, if Charlie's measurement result is 1 with probability $p_{1}$, then
the system state after his measurement is $|\Omega_{1}\rangle_{ab}|\phi^{+}\rangle_{a^{'}b^{'}}$.
Using the Schmidt-Decomposition we get
$$|\Omega_{1}\rangle_{ab}|\phi^{+}\rangle_{a^{'}b^{'}}
=\frac{1}{\sqrt{2}}{(\sqrt{\lambda_{10}}|\overline{0}\,\overline{0}}\rangle
+\sqrt{\lambda_{11}}|\overline{1}\,\overline{1}\rangle)_{ab}(|00\rangle+|11\rangle)_{a^{'}b^{'}},$$
where $\lambda_{10}$ and $\lambda_{11}$ are the same as those in section 2.
Bob can also reconstruct the two-qubit state using similar method in the above three steps. As a result, for the real case, the total successful probability for the sender Alice to prepare the two-qubit state at the position of Bob under the control of controller Charlie is
\begin{eqnarray*}
&&2\times [p_{0}\frac{\lambda_{00}(\alpha^{2}+\beta^{2})
+\lambda_{01}(\gamma^{2}+\delta^{2})}{2}\frac{\lambda_{00}}{\lambda_{00}(\alpha^{2}+\beta^{2})
+\lambda_{01}(\gamma^{2}+\delta^{2})}
\nonumber\\&&+p_{0}\frac{\lambda_{00}(\gamma^{2}+\delta^{2})+\lambda_{01}(\alpha^{2}
+\beta^{2})}{2}\frac{\lambda_{00}}{\lambda_{00}(\gamma^{2}+\delta^{2})+\lambda_{01}(\alpha^{2}+\beta^{2})}
\nonumber\\&&+p_{1}\frac{\lambda_{10}(\alpha^{2}+\beta^{2})
+\lambda_{11}(\gamma^{2}+\delta^{2})}{2}\frac{\lambda_{10}}{\lambda_{10}(\alpha^{2}+\beta^{2})
+\lambda_{11}(\gamma^{2}+\delta^{2})}
\nonumber\\&&+p_{1}\frac{\lambda_{10}(\gamma^{2}+\delta^{2})+\lambda_{11}(\alpha^{2}
+\beta^{2})}{2}\frac{\lambda_{10}}{\lambda_{10}(\gamma^{2}+\delta^{2})+\lambda_{11}(\alpha^{2}+\beta^{2})}]
\nonumber\\&&=2(p_{0}\lambda_{00}+p_{1}\lambda_{10}).
\end{eqnarray*}
It is the same as that of the controlled teleportation for the real case of a qubit. In the whole
process the consumption of classical resource is 3 cbits.

For the case in which there is at least one complex coefficient, in \emph{Step 2}, Alice constructs measurement basis$\{|\nu_{0}\rangle, |\nu_{1}\rangle, |\nu_{2}\rangle, |\nu_{3}\rangle\}_{aa^{'}}$ in the following form,
$$\left(
  \begin{array}{c}
    |\nu_{0}\rangle \\
    |\nu_{1}\rangle \\
    |\nu_{2}\rangle \\
    |\nu_{3}\rangle \\
  \end{array}
\right)_{aa^{'}}=\left(
          \begin{array}{cccc}
            \alpha^{*} & -\beta^{*}   & \gamma^{*}  & -\delta^{*} \\
            \zeta\alpha^{*}  & -\zeta\beta^{*} & -\zeta^{-1}\gamma^{*} & \zeta^{-1}\delta^{*} \\
            -\beta & -\alpha  & -\delta & -\gamma \\
            -\zeta\beta & -\zeta\alpha & \zeta^{-1}\delta   & \zeta^{-1}\gamma \\
          \end{array}
        \right)\left(
                 \begin{array}{c}
                   |0^{'}0\rangle \\
                   |0^{'}1\rangle \\
                   |1^{'}0\rangle \\
                   |1^{'}1\rangle \\
                 \end{array}
               \right)_{aa^{'}},
$$
where $\zeta=\sqrt{(|\gamma|^{2}+|\delta|^{2})/(|\alpha|^{2}+|\beta|^{2})}$, here we can assume that $|\alpha|^{2}+|\beta|^{2} \neq 0$. Because if $|\alpha|^{2}+|\beta|^{2}=0$, the number of coefficients decrease to two, which is actually the same as the single qubit case. The system for Alice and Bob can be reexpressed as 
\begin{eqnarray}
&&|\Omega_{0}\rangle_{ab}|\phi^{+}\rangle_{a^{'}b^{'}}
\nonumber\\&=&\frac{1}{\sqrt{2}}\bigg \{|\nu_{0}\rangle[\sqrt{\lambda_{00}}(\alpha|0^{'}0\rangle
-\beta|0^{'}1\rangle)+\sqrt{\lambda_{01}}(\gamma|1^{'}0\rangle
-\delta|1^{'}1\rangle)]
\nonumber\\&&+|\nu_{1}\rangle[\sqrt{\lambda_{00}}\zeta(\alpha|0^{'}0\rangle
-\beta|0^{'}1\rangle)-\sqrt{\lambda_{01}}\zeta^{-1}(\gamma|1^{'}0\rangle
-\delta|1^{'}1\rangle)]
\nonumber\\&&-|\nu_{2}\rangle[\sqrt{\lambda_{00}}(\beta^{*}|0^{'}0\rangle
+\alpha^{*}|0^{'}1\rangle)+\sqrt{\lambda_{01}}(\delta^{*}|1^{'}0\rangle
+\gamma^{*}|1^{'}1\rangle)]
\nonumber\\&&+|\nu_{3}\rangle[\sqrt{\lambda_{00}}\zeta(-\beta^{*}|0^{'}0\rangle
-\alpha^{*}|0^{'}1\rangle)+\sqrt{\lambda_{01}}\zeta^{-1}(\delta^{*}|1^{'}0\rangle
+\gamma^{*}|1^{'}1\rangle)]\bigg\}_{aa^{'}bb^{'}}.
\end{eqnarray}
Thus Alice can get result 0 and 1 with probability 
$[\lambda_{00}(|\alpha|^{2}+|\beta|^{2})+\lambda_{01}(|\gamma|^{2}+|\delta|^{2})]/2$ and 
$[\lambda_{00}\zeta^{2}(|\alpha|^{2}+|\beta|^{2})+\lambda_{01}\zeta^{-2}(|\gamma|^{2}+|\delta|^{2})]/2$, respectively. The states after Alice's measurement with respect to the result 0 and 1 are 
\begin{eqnarray}
|\vartheta_{0}\rangle=\frac{\sqrt{\lambda_{00}}(\alpha|0^{'}0\rangle
-\beta|0^{'}1\rangle)+ \sqrt{\lambda_{01}}(\gamma|1^{'}0\rangle
-\delta|1^{'}1\rangle)}{\sqrt{\lambda_{00}(|\alpha|^{2}+|\beta|^{2})
+\lambda_{01}(|\gamma|^{2}+|\delta|^{2})}}
\end{eqnarray}
and 
\begin{eqnarray}
|\vartheta_{1}\rangle=\frac{\sqrt{\lambda_{00}}\zeta(\alpha|0^{'}0\rangle
-\beta|0^{'}1\rangle)+ \sqrt{\lambda_{01}}\zeta^{-1}(-\gamma|1^{'}0\rangle
+\delta|1^{'}1\rangle)}{\sqrt{\lambda_{00}\zeta^{2}(|\alpha|^{2}+|\beta|^{2})
+\lambda_{01}\zeta^{-2}(|\gamma|^{2}+|\delta|^{2})}}
\end{eqnarray}
We divide into two cases according to the value of $\zeta$.\\
(i) $\zeta=1$, i.e. $|\alpha|^{2}+|\beta|^{2}=|\gamma|^{2}+|\delta|^{2}$.
In this case, using similar methods as in the real cases above, Bob can recover the desired two-qubit state both from states in Eq.(11) and Eq.(12). And the probabilities are both 
$\lambda_{00}/(\lambda_{00}(|\alpha|^{2}+|\beta|^{2})+\lambda_{01}(|\gamma|^{2}+|\delta|^{2}))$.
Similar scheme applies to the case that Charlie's measurement result is 1. Thus the total successful
probability for Alice remotely to prepare the two-qubit state $|\varphi\rangle$ at Bob's position under
the control of Charlie is 
\begin{eqnarray*}
&&2\times \bigg\{p_{0}[\frac{\lambda_{00}(|\alpha|^{2}+|\beta|^{2})+\lambda_{01}(|\gamma|^{2}+|\delta|^{2})}{2}]
\frac{\lambda_{00}}{\lambda_{00}(|\alpha|^{2}+|\beta|^{2})+\lambda_{01}(|\gamma|^{2}+|\delta|^{2})}
\nonumber\\&&
+p_{1}[\frac{\lambda_{10}(|\alpha|^{2}+|\beta|^{2})+\lambda_{11}(|\gamma|^{2}+|\delta|^{2})}{2}]
\frac{\lambda_{10}}{\lambda_{10}(|\alpha|^{2}+|\beta|^{2})+\lambda_{11}(|\gamma|^{2}+|\delta|^{2})}\bigg\}  
\nonumber\\&&=p_{0}\lambda_{00}+p_{1}\lambda_{10}, 
\end{eqnarray*}
which is half of the case that all the coefficients are real. As for the result 3 and 4, the CRSP protocol fails.\\
(ii) $\zeta\neq 1$. For this case, as Bob does not know the classical information of $|\varphi\rangle$, only when Alice's measurement result is 0, Bob can reconstruct the two-qubit state $|\varphi\rangle$. Thus the successful probability reduces to half of (i) as $(p_{0}\lambda_{00}+p_{1}\lambda_{10})/2$.  

\section{Conclusions}
In this paper, protocols for controlled remote state preparation are presented both for a single qubit and two-qubit state.
We utilize the general pure three qubit states as the shared quantum channels, which are not LOCC equivalent to the GHZ state. 
We discuss protocols for both states with real and complex coefficients, and find that the general pure three-qubit states can help to complete CRSP probabilistically. More than that, in some spacial cases, the CRSP can be achieved with
unit probability, which are deterministic CRSP protocols. This overcomes the limitation that most of the existing quantum 
communication protocols are completed with GHZ-, W- or Bell states, or the composition of these states. Moreover, due to the 
involvement of controller and multi-partities, this work may have potential application in controlled quantum communication, 
quantum network communication and distributed computation.

%\label{sec:1}
%Text with citations \cite{RefB} and \cite{RefJ}.
%\subsection{Subsection title}
%\label{sec:2}
%as required. Don't forget to give each section
%and subsection a unique label (see Sect.~\ref{sec:1}).
%\paragraph{Paragraph headings} Use paragraph headings as needed.
%\begin{equation}
%\end{equation}

%\begin{figure}
%  \includegraphics{example.eps}
%\end{figure}
%\begin{figure*}
%  \includegraphics[width=0.75\textwidth]{example.eps}
%\end{figure*}
%\begin{table}
%\begin{tabular}{lll}
%\hline\noalign{\smallskip}
%\end{tabular}
%\end{table}

\begin{acknowledgements}
We thank the anonymous reviewers for valuable comments.
\end{acknowledgements}


\begin{thebibliography}{}
\bibitem{Ben1} Bennett, C.H., Brassard, G.,  Cr\'{e}peau, C.,  Jozsa, R.,  Peres, A., Wootters, W.K.: Teleporting
an unknown quantum state via dual classical and Einstein-Podolsky-Rosenchannels. Phys. Rev. Lett. \textbf{70}, 1895-1899 (1993)

\bibitem{Lee1} Lee, H.W., Kim, J.: Quantum teleportation and Bell's inequality using single-particle entanglement.
 Phys. Rev. A \textbf{63}, 012305 (2001)

\bibitem{Kim}  Kim, Y.H., Kulik, S.P., Shih, Y.: Quantum teleportation of a polarization state with a complete Bell state measurement.
Phys. Rev. Lett. \textbf{86}, 1370-1373 (2001)

\bibitem{Zhou} Zhou, J.-D., Hou, G, Zhang, Y.-D.: Teleportation scheme of S-level quantum pure states by two-level
Einstein-Podolsky-Rosen states. Phys. Rev. A \textbf{64}, 012301 (2001)

\bibitem{Wang1} Wang, X.-G.: Quantum teleportation of entangled coherent states. Phys. Rev. A \textbf{64}, 022302, (2001)

\bibitem{van} van Enk, S.J., Hirota, O.: Entangled coherent states: Teleportation and decoherence. Phys. Rev. A \textbf{64}, 022313 (2001)

\bibitem{Bou} Bouwmeester, D. Pan J.-W. , Mattle, K., \emph{et al.}: Experimental quantum teleportation. Nature \textbf{390}, 575 (1997)

\bibitem{Bra} Braunstein, S. L., Kimble, H.: Teleportation of continuous quantum variables. Phys. Rev. Lett. \textbf{80}, 869-872 (1998)

\bibitem{Fur} Furusawa, A. Srensen, J. L., Braunstein, S. L., \emph{et al.}: Unconditional quantum teleportation. Science, \textbf{282}, 706-709 (1998)

\bibitem{Jin} Jin, X.-M., Ren, J.-G, Yang, B., et al.: Experimental free-space quantum teleportation. Nature Photonics \textbf{4}, 376-381 (2010)

\bibitem{Hua} Huang, Y.-F., Ren, X.-F., Zhang, Y.-S., Duan, L.-M., Guo, G.-C.:  Experimental teleportation of a quantum controlled-NOT gate. Phys. Rev. Lett. \textbf{93}, 240501 (2004)

\bibitem{Noe} Christian, N., Andreas, N., Andreas, R., \emph{et al.}: Efficient teleportation between remote single-atom quantum memories. Phys. Rev. Lett. \textbf{110}, 140403 (2013)

\bibitem{Nil} Nilsson, J., Stevenson, R. M., Chan, K. H. A., \emph{et al.}: Quantum teleportation using a light-emitting diode.
Nature Photonics \textbf{7}, 311-315 (2013)

\bibitem{Fri} Friis, N., Lee, A. R., Truong, K., \emph{et al.}: Relativistic quantum teleportation with superconducting circuits.
 Phys. Rev. Lett. \textbf{110}, 113602 (2013)

\bibitem{Lo} Lo, H. K.: Classical-communication cost in distributed quantum-information processing: A generalization of quantum-communication complexity. Phys. Rev. A \textbf{62}, 012313 (2000)

\bibitem{Ben2} Bennett, C. H., Divincenzo, D. P., Shor, P. W., Smolin, J. A., Terhal, B. M., Wootters, W. K.:  Remote state preparation. Phys. Rev. Lett. \textbf{87}, 077902 (2001)

\bibitem{Dev} Devetak, I., Berger, T.: Low-entanglement remote state preparation. Phys. Rev. Lett. \textbf{87}, 197901 (2001)

\bibitem{Guo} Zheng, Y.-Z., Gu, Y.-J., Guo, G.-C.: Remote state preparation via a non-maximally entangled channel.
Chin. Phys. Lett. \textbf{19}, 14-16 (2002)

\bibitem{Zeng}Zeng, B, Zhang, P.: Remote-state preparation in higher dimension and the parallelizable manifold $S^{n-1}$.
 Phys. Rev. A \textbf{65}, 022316 (2002)

\bibitem{Shi} Shi, B.-S., Tomita, A.: Remote state preparation of an entangled state.
J. Phys. B: At. Mol. Opt. Phys. \textbf{4}, 380-382, (2002)

\bibitem{Berry} Berry, D.-W., Sanders, B.-C.: Optimal remote state preparation. Phys. Rev. Lett. \textbf{90}, 057901 (2003)

\bibitem{Leung} Leung, D.-W., Shor, P.-W.: Oblivious remote state preparation. Phys. Rev. Lett. \textbf{90}, 127905 (2003)

\bibitem{Hayashi} Hayashi, A., Hashimoto, T., Horibe, M.: Remote state preparation without
oblivious conditions. Phys. Rev. A \textbf{67}, 052302 (2003)

\bibitem{Yu} Yu, C., Song, H., Wang, Y.: Remote preparation of a qudit using maximally entangled
states of qubits. Phys. Rev. A \textbf{73}, 022340 (2006)

\bibitem{Kur} Kurucz, Z., Adam, P., Kis, Z., Janszky, J.: Continuous variable remote state
preparation. Phys. Rev. A \textbf{72}, 052315 (2005)

\bibitem{Lee2} Lee, S.: Bound on remote preparation of entanglement from isotropic states.
Phys. Rev. A \textbf{85}, 052311 (2012)

\bibitem{Peng} Peng, X.-H., Zhu, X.-W., Fang, X.-M., \emph{et al.}: Experimental implementation of
remote state preparation by nuclear magnetic resonance. Phys. Lett. A \textbf{306}, 271-276 (2003)

\bibitem{Bar} Barreiro, J. T., Wei, T.-C., Kwiat, P. G., Remote preparation of single-photon
'Hybrid' entangled and vector-polarization states. Phys. Rev. Lett. \textbf{105}, 3 (2010)

\bibitem{WangZY1} Wang, Z.-Y., Liu Y.-M., Zuo X.-Q.: Controlled remote state preparation.
Commun. Theor. Phys. \textbf{52}, 235-240 (2009)

\bibitem{HouK} Hou, K., Wang J., Yuan H.: Multiparty-controlled remote preparation of two-particle state.
Commun. Theor. Phys. \textbf{52}, 848-852 (2009)

\bibitem{WangZY2} Wang, Z.-Y.: Controlled remote preparation of a two-qubit state via an asymmetric quantum channel.
Commun. Theor. Phys. \textbf{55} 244-250 (2011)

\bibitem{WangZY3} Song, J.-F., Wang, Z.-Y.:Controlled remote preparation of a two-qubit state
via positive operator-valued measure and two three-qubit entanglements. Int. J. Theor. Phys. \textbf{50} 2410-2425 (2011)

\bibitem{LiZ} Li, Z., Zhou, P.: Probabilistic Multiparty-controlled remote preparation of an arbitrary
m-qudit state via positive operator-valued measurement. Int. J. Quantum Inf. \textbf{10} 1250062 (2012)

\bibitem{ChenXB} Chen, X.-B., Ma, S.-Y., Su, Y.:
Controlled remote state preparation of arbitrary two and three qubit states via the Brown state.
Quantum. Inf. Process. \textbf{11} 1653-1667 (2012)

\bibitem{AAcin} Ac\'{\i}n, A., Andrianov, A., Costa, L., Jan\'{e}, E., Latorre, J. I., Tarrach,R.:
Generalised Schmidt decomposition and classification of three-quantum-bit states. Phys. Rev. Lett. \textbf{85} 1560 (2000)

\bibitem{GaoT} Gao, T., Yan, F.-L., Li, Y.-C.: Optimal controlled teleportation. EPL, \textbf{84} 50001 (2008)
% etc
\end{thebibliography}


\end{document}